\documentclass[10pt,preprintnumbers,amsmath,amssymb,floatfix,twocolumn,prx]{revtex4-1}
\usepackage[utf8]{inputenc}
\usepackage{amsmath}
\usepackage{mathrsfs}
\usepackage{graphics}
\usepackage{xspace}
\usepackage[T1]{fontenc}
\usepackage{graphicx}% Include figure files
\usepackage{dcolumn}% Align table columns on decimal point
\usepackage{bm}% bold math
\usepackage{float}%
\usepackage{color}

\begin{document}

\title{Spin-state ice in geometrically frustrated spin-crossover materials}
\author{Jace Cruddas}
\email{j.cruddas@uq.edu.au}
\affiliation{School of Mathematics and Physics, The University of Queensland, QLD 4072, Australia}
\author{B. J. Powell}
\affiliation{School of Mathematics and Physics, The University of Queensland, QLD 4072, Australia}

\begin{abstract}
	Spin crossover materials contain metal ions that can access two spin-states: one low-spin (LS), the other high-spin (HS). We propose that  frustrated elastic interactions  can give rise to  spin-state ices -- phases of matter without long-range order,  characterized by a local constraint or `ice rule'. The low-energy physics of spin-state ices is described by an emergent divergence-less gauge field with a gap to topological excitations that are deconfined quasi-particles with spin fractionalized midway between the spins of the LS and HS states.
\end{abstract}

\maketitle

\section{Introduction}

\begin{figure}
	\centering
	\includegraphics[width=0.9\columnwidth]{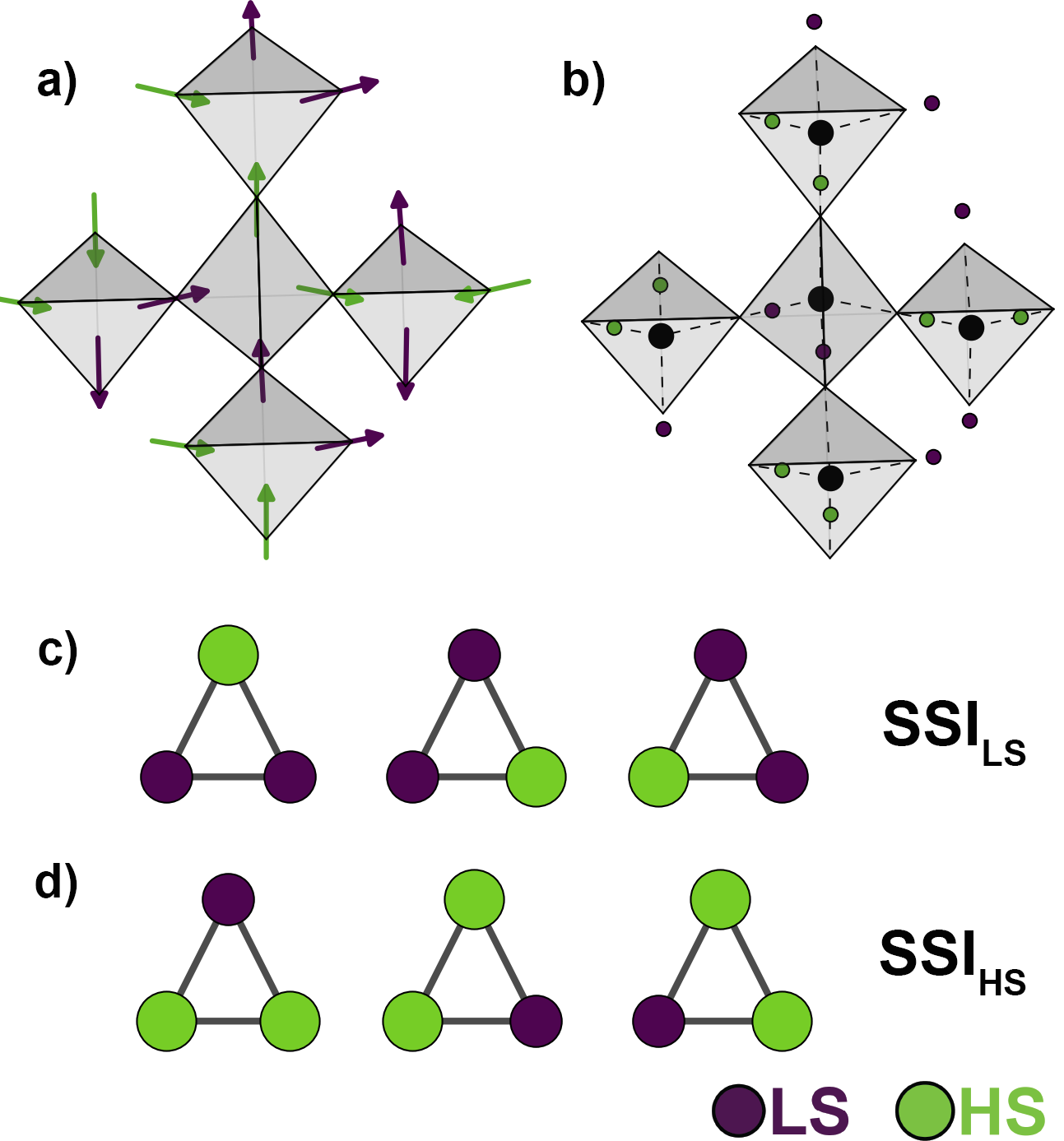} 	
	\caption{
		Ice rules in (a) spin ice, (b) water ice, and (c,d) spin-state ice (SSI). The ice rules for these systems are: (a) In spin ice two spins must point into each tetrahedron and the other two must point out. (b) In water ice each oxygen must form short (covalent) bonds with two hydrogens atoms and long (hydrogen) bonds with another two. (c) In SSI$_\text{LS}$ each triangle of the kagome lattice must contain two low spin (LS) metal ions and one high spin (HS) ion. (d) In SSI$_\text{HS}$ each triangle of the kagome lattice must contain two HS ions and one LS ion.
	}
	\label{Fig:rules}  
\end{figure}

Frustration, the inability to simultaneously minimize competing interactions, can  produce macroscopically degenerate classical states \cite{Pauling,lieb,IsingKagome} and long-range entangled  quantum states \cite{Wen},  notably topological spin-liquids \cite{Balents,CastelnovoARCMP}.   Spin ices, such as Dy$_2$Ti$_2$O$_7$ and Ho$_2$Ti$_2$O$_7$, are an important class of frustrated magnets. In these systems the magnetic atoms form a pyrochlore lattice, composed of vertex sharing tetrahedra, Fig. \ref{Fig:rules}a. The interactions between the magnetic moments are minimized if two spins point into each tetrahedron and the other two spins  point out. A macroscopic number of states  satisfy this `ice rule', giving rise to an extensive residual entropy. Because the ice rule is a local constraint the low-energy physics of spin ices is described by an emergent divergence-less gauge field \cite{CastelnovoARCMP,Henley}, which gives rise to pinch point singularities in neutron scattering structure factor \cite{Fennell}. The quasiparticle excitations above these degenerate states are magnetic monopoles \cite{CastelnovoNature}, which are necessarily fractionalized excitations as magnetic monopoles have not been observed in the vacuum.

%Spin crossover materials contain molecules with two thermodynamically accessible spin-states: one low-spin (LS) and the other high-spin (HS) \cite{Kahn,Goodwin,Halcrow}. Here we show that  frustrated elastic interactions  can give rise to  spin-state ices. The low-energy physics exhibits an emergent divergence-less field with a gap to topological excitations that are deconfined quasi-particles with spin fractionalized midway between the LS and HS spin.

The prototypical example of ice physics is water ice, Fig. \ref{Fig:rules}b. The oxygen ions form a regular lattice with a proton between each pair of oxygen ions; forming a covalent bond with one and a hydrogen bond with the other. Remarkably, the protons lack any kind of long-range order. Any configuration, obeying the ice rules, of two covalent bonds and two hydrogen bonds per oxygen atom is degenerate \cite{BernalFowler}. This leads to a macroscopic residual ($T=0$) entropy \cite{Pauling,lieb}. But it is not known how commonly ice physics occurs in other classes of materials. 

In this paper we propose that ice physics can be realized by frustrated elastic interactions in spin crossover (SCO) materials. Metal ions in SCO materials can access two different spin-states: one low-spin (LS) and the other high-spin (HS) \cite{Kahn,Goodwin,Halcrow}. In `spin-state ice' the spin-states of the metal ions are disordered, but follow a local ice rule, Fig. \ref{Fig:rules}c,d. The low-energy physics is therefore described by a gauge field. We demonstrate this by showing that the pseudospin structure factor displays pinch point singularities, characteristic of the Coulomb phase \cite{Henley}. Importantly, we also show that the spin structure factor in a weak magnetic field is proportional to the pseudospin structure factor -- therefore, neutron scattering experiments provide a potential smoking gun experiment for identifying spin-state ices. We also show that the low-energy excitations are fractionalized, with spins midway between the spins of the HS and LS states. Finally, we argue that this physics is responsible for the disordered spin-states observed  \cite{Sekine} in kagome lattice [Co$_2$Fe$_2$] SCO complexes.

\begin{figure}
	\centering
	\includegraphics[width=\columnwidth]{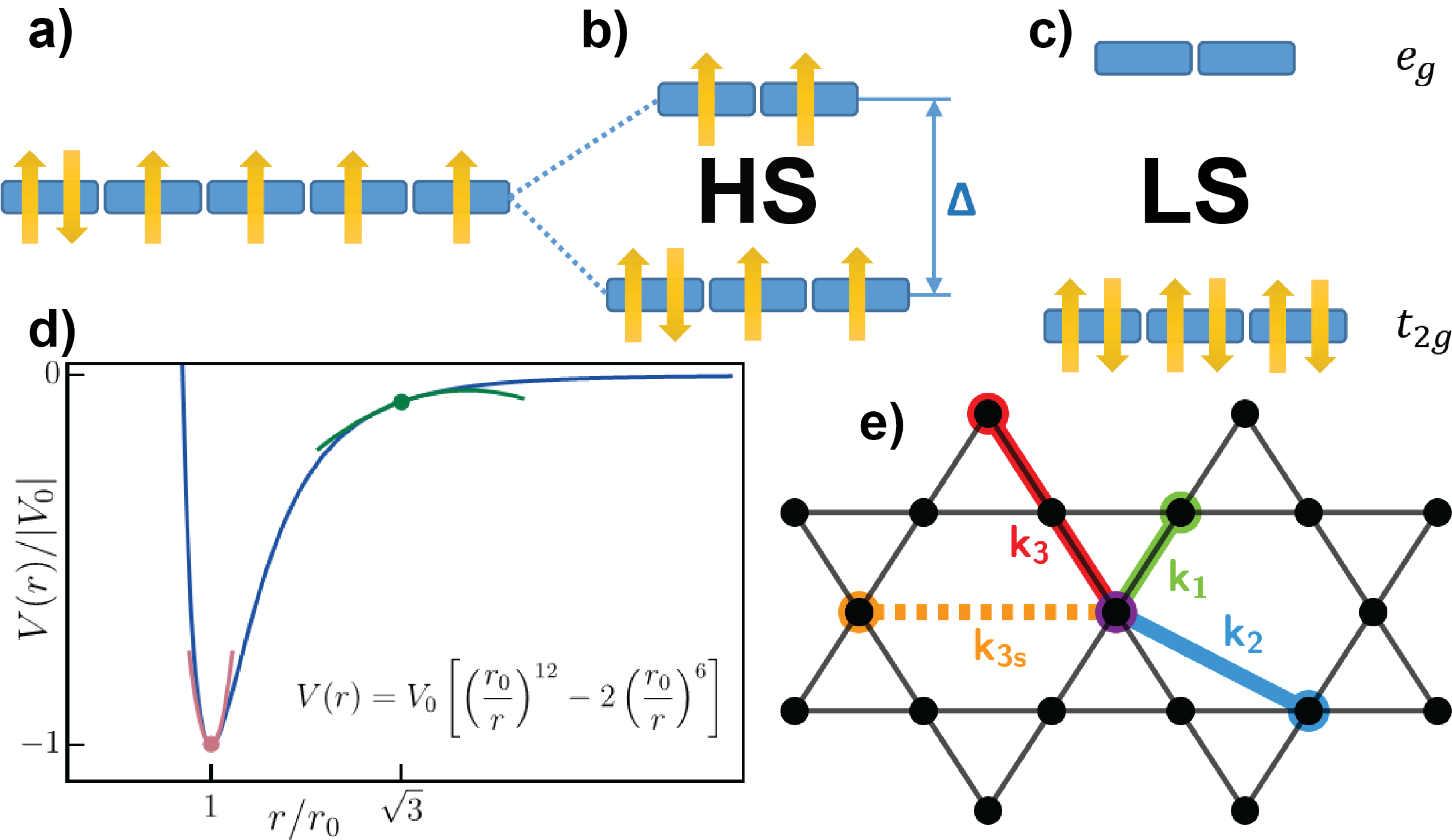} 	
	\caption{
		Outline of the model of spin-crossover materials. a) Six electrons in five d-orbitals of a transition metal, such as Fe$^{2+}$. When placed in an organic cage the orbitals are split into $e_g$ and $t_{2g}$ levels, separated by an energy, $\Delta$. 
		b) The high-spin (HS) state is filled according to Hund’s first rule. 
		c) For shorter bond lengths $\Delta$ increases and the orbitals are filled according to the \textit{aufbau} principle resulting in the low-spin (LS) state. 
		d) The Lennard-Jones potential, $V(r)$ (blue curve) between two molecules separated by a distance $r$. Near the minimum, $V(r_0)=V_0$, the second derivative is positive (pink curve), whereas at larger distances the second derivative becomes negative (green curve). Thus, if the nearest neighbor separation $x\simeq r_0$ we expect $k_1>0$ but $k_2<0$ and $k_3<0$. Note that in general away from the minimum there are also linear terms and higher order terms, which, because of the local $Z_2$ symmetry of the model,  simply renormalize $H_\text{eff}$ and  $J_n$. 
		e) The kagome lattice, with the three nearest neighbor interactions, $k_1$-$k_3$, marked. The through space interaction, $k_{3s}$ is not included in our model.
	}
	\label{Fig:model}  
\end{figure}

SCO is typically observed in coordination complexes and frameworks, these materials contain metal ions (often Fe$^\text{II}$)  surrounded by several ligands. 
The five $d$-orbitals of a single $\text{Fe}^{2+}$ ion contain six electrons, Fig.    \ref{Fig:model}a, with a ground state given by Hund’s rules. However, in a  complex or framework the $d$-orbitals of an  $\text{Fe}^\text{II}$ ion are split by the ligand field, $\Delta$, into, say, $t_{2g}$ and $e_g$ orbitals, Fig. \ref{Fig:model}b,c. 
If this splitting is small compared to the Hund’s rule coupling then the ground state is HS, Hund’s first rule is obeyed, Fig. \ref{Fig:model}b. However, if the splitting is large then the ground state is LS, in accordance with the \textit{aufbau} principal, Fig. \ref{Fig:model}c. In general there will be a enthalpy difference between the HS and LS states  $\varepsilon=\varepsilon_{HS}-\varepsilon_{LS}$, where $\varepsilon_{x}$ is the enthalpy of a single molecule in  spin-state $x$.

The $t_{2g}$ orbitals are bonding and the $e_g$ orbitals are antibonding. Thus, the metal-ligand bond length in the HS state is significantly larger than that in the LS state. This increase  is often as much as $10$~\% \cite{Kahn,Goodwin,Halcrow}.   

Molecules have higher entropies in their HS states than in their LS states. The most obvious source of this entropy difference is the electronic (spin) entropy, $S_{e}=k_B\ln(2S+1)$. For Fe$^\text{II}$ the HS state has spin $S=2$ and the LS state is $S=0$, so $\Delta S_e=S_e^\text{HS}-S_e^\text{LS}=k_B\ln5$, where $S_e^{x}$ is electronic contribution to the entropy of spin-state $x$. However, the two spin-states can also have different orbital entropies and the lengthening of the metal-ligand bonds concomitant with the change of spin-state also softens the associated vibrational modes in the HS complex, further increasing the entropy difference,  $\Delta S$. Typically the vibrational contribution is largest and $ \Delta S\sim4\Delta S_e $ \cite{Goodwin,Sorai1,Sorai2}.

\section{Psuedospin model and spin-state ice at zero temperature}

To describe the many-body physics of SCO materials it is convenient to use a pseudospin notation where the spin-state of each metal center is  labeled by a binary variable  $\sigma_i$, we take $\sigma_i=1$ if the $i^\text{th}$ metal center is HS and $\sigma_i=-1$ if it is LS. As magnetic interactions between the spins are weak in most SCO materials, it is convenient to absorb the entropy of a single cluster into the Hamiltonian \cite{WP}. This has the side effect of making the effective Hamiltonian for a single molecule appear temperature dependent:
\begin{eqnarray}
\mathcal{H}_0=\sum_i \frac12 \left( \varepsilon - T\Delta S \right) \sigma_i \equiv H_\text{eff} \sum_i \sigma_i.
\label{eq:Heff}
\end{eqnarray}
 
Clearly $\varepsilon<0$ favors majority HS states at all temperatures. Similarly if $\varepsilon$ is large and positive majority LS states arise. Both cases are commonly observed and correspond to single molecule magnets and non-magnetic molecules respectively. However, at $T=\varepsilon/\Delta S$ one expects a crossover between the low-enthalpy, low-entropy LS state and the high-enthalpy, high-entropy HS state. This SCO is indeed observed in many materials and these form a large and active field of chemistry with hundreds of known examples \cite{Goodwin,Halcrow}. Importantly, in the solid state, significant hysteresis is often found indicating a first order transition and thus non-trivial interactions between metal centers. Furthermore, multi-step transitions are also observed. Often the intermediate states display ordered patterns of HS and LS metal centers reminiscent of antiferromagnetic order \cite{Murphy,Clements,Griffin,Collet}.

\begin{figure}
	\includegraphics[width=\columnwidth]{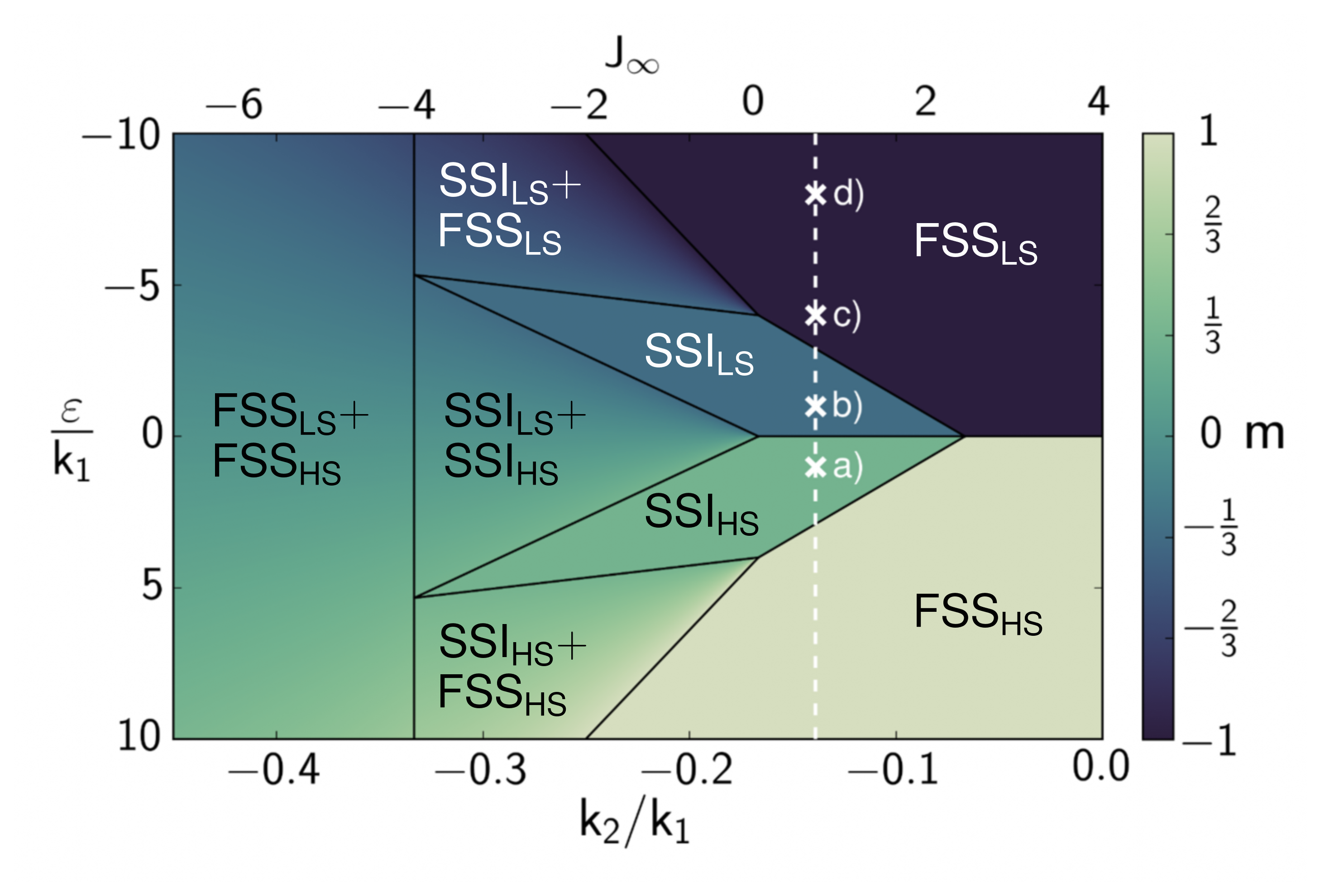}  
	\centering
	\caption{
		The zero temperature phase diagram for model (\ref{model}) of spin-crossover materials on the kagome lattice for $k_1>0$ and $k_3=(3/4) k_2$. SSI indicates spin-state ice, FSS is ferro-spin-state; subscripts indicates the majority spin-state. The coexistance regions are driven by the interplay of the short-range elastic interactions, which prefer either SSI ($k_2>-k_1/3$) or FSS ($k_2<-k_1/3$) order, and the long-range strain ($J_\infty$), which favors minimizing (maximizing) the pseudospin magnetization for $J_\infty<0$ ($J_\infty>0$). Thus, $J_\infty$ drives the formation of defects, which  cluster together to minimize the short range interactions. The vertical dashed white line and the points labeled a-d show the parameters explored in Fig. \ref{Fig:thermo}.
	}\label{fig:phase-dia}
\end{figure}

In a crystal the change in size of a molecule undergoing a spin-state transition induces elastic interactions on neighboring molecules \cite{review,Nishino}. 
The simplest model \cite{Paez} of such interactions is via a spring constant, $k_n$, between $n^{th}$ nearest neighbor molecules
\begin{eqnarray}\label{Paez}
\mathcal{H}_n&=&\frac{k_n}{2}\sum_{\langle i,j\rangle_n}\left\{r_{i,j}-[R(\sigma_i)+R(\sigma_j)]  \right\}^2,
\label{Ham:elastic}
\end{eqnarray}
where $R(\sigma_i)$ is the radius of the $i^\text{th}$ molecule in spin-state $\sigma_i$ such that $R(1)>R(-1)$, $r_{i,j}$ is the instantaneous bond distance between sites $i$ and $j$, and $\langle i,j\rangle_n$ indicates that the sum runs over all $n^\text{th}$ nearest neighbors.
Models including these interactions have been shown to reproduce thermal hysteresis and multi-step SCO transitions \cite{Nishino,Paez}. 

\begin{figure*}
	\centering
	\includegraphics[width=\textwidth]{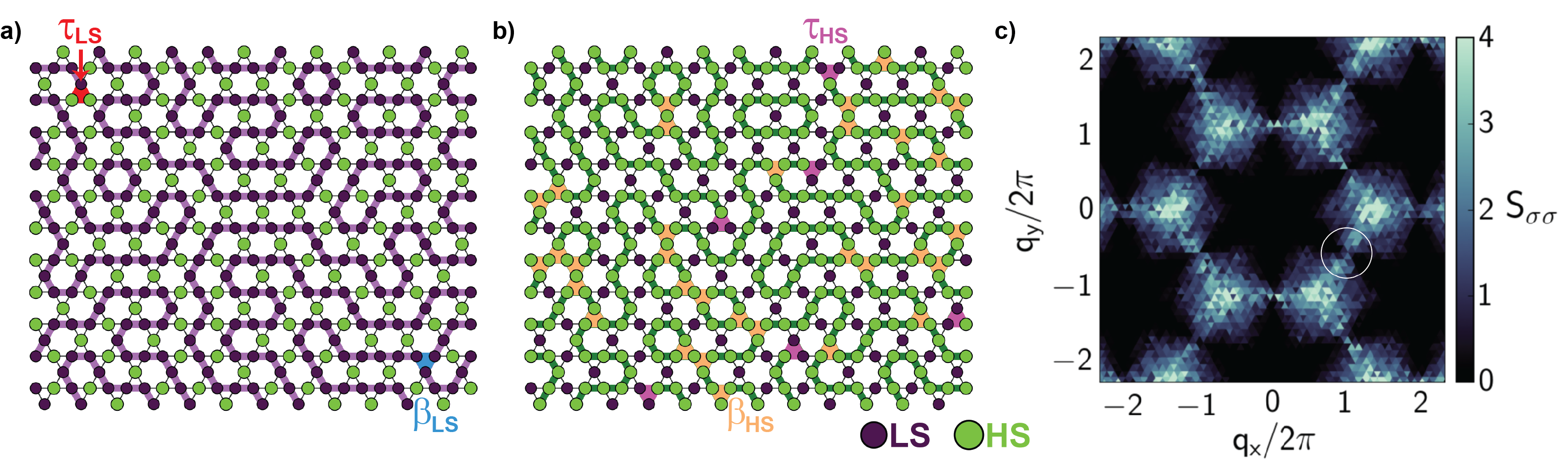}
	\caption{
		Spin state ice at $T>0$ with defects and strings of the majority spin-state highlighted. 
		a) Typical snapshot of a region of spin-state ice (SSI$_\text{LS}$). Almost every triangle has two low-spin and one high-spin metal centers, obeying the spin-state ice rules. There are two exceptions: $\beta_{LS}$ a triangle with three LS metal centers, which leads to a branching of a string of LS metal centers, and $\tau_{LS}$ a triangle with two LS and one HS metal centers, which causes a string of LS metal centers to terminate. 
		b) Snapshot of a region of SSI$_\text{HS}$ as it begins to melt in the effective magnetic field, $H_\text{eff}$. Most triangles still obey the spin-state ice rule, but there significant numbers of $\beta_{HS}$ and $\tau_{HS}$ defects. There are more $\beta_{HS}$ than $\tau_{HS}$ defects leading to a fraction of HS metal centers, $n_{HS}=0.710$ for the configuration this snapshot is taken from, considerably greater than in defectless SSI$_\text{HS}$, where $n_{HS}=2/3$. Note also the tendency for the $\beta_{HS}$ defects to cluster. There are no triangles containing three LS metal centers, this is typical. The full lattices from which these regions are taken are shown in Figs. S1 and S2 \cite{SI}. 
		c) Structure factor for spin ice. 
		The pseudospin structure factor, ${\mathcal S}_{\sigma\sigma}$, cf. Eq. (\ref{eq:structure-factor-pseduospin}), displays pinch points at the Brillouin zone boundaries characteristic of spin-state ice, one of which is circled. This indicates that the low-energy physics of spin-state ice is described by a divergence-less gauge field \cite{CastelnovoARCMP,Henley}. In a magnetic field sufficient to polarize the spins the spin structure factor, ${\mathcal S}_{zz}$, is proportional to the pseudospin structure factor, cf. Eq. (\ref{eq:structure-factor-realspin}). Thus, the pinch points are directly detectable via neutron scattering. 
		Snapshots taken at (a) $k_BT=0.111 k_1$ and $k_2=-0.139k_1$ and (b) $k_BT=1.32 k_1$ and $k_2=-0.05k_1$; in both  $\varepsilon=1.5k_1$, $k_3=(3/4)k_2$ and $\Delta S=4k_B\ln5$.
		Structure factor calculated for $k_2/k_1=-1.39$, $k_3=(3/4)k_2$, $k_BT=0.11k_1$ and $H_\text{eff}=-1.32 k_1$.
	}\label{Fig:snapshots}
\end{figure*}

\begin{figure*}
%	(a) $\varepsilon=-k_1$ \hspace{0.5\columnwidth} (b) $\varepsilon=k_1$ \hspace{0.5\columnwidth} (c) $\varepsilon=4k_1$  \\
%	\hspace{0.5\columnwidth}
%	\includegraphics[width=0.67\columnwidth]{1_yeahbuddy}\vspace{5pt}
%	\includegraphics[width=0.67\columnwidth]{4_yeahbuddy_2}   
%	\\	(d) $\varepsilon=8k_1$ \hspace*{2cm} \hspace{0.5\columnwidth} (e) $\varepsilon=12k_1$ \hspace{0.5\columnwidth} (f) $\varepsilon=16k_1$ \\ 
%	\includegraphics[width=0.67\columnwidth]{8_yeahbuddy_2}
%	\includegraphics[width=0.67\columnwidth]{12_yeahbuddy_2}  
%	\includegraphics[width=0.67\columnwidth]{16_yeahbuddy_2}  
	\includegraphics[width=\textwidth]{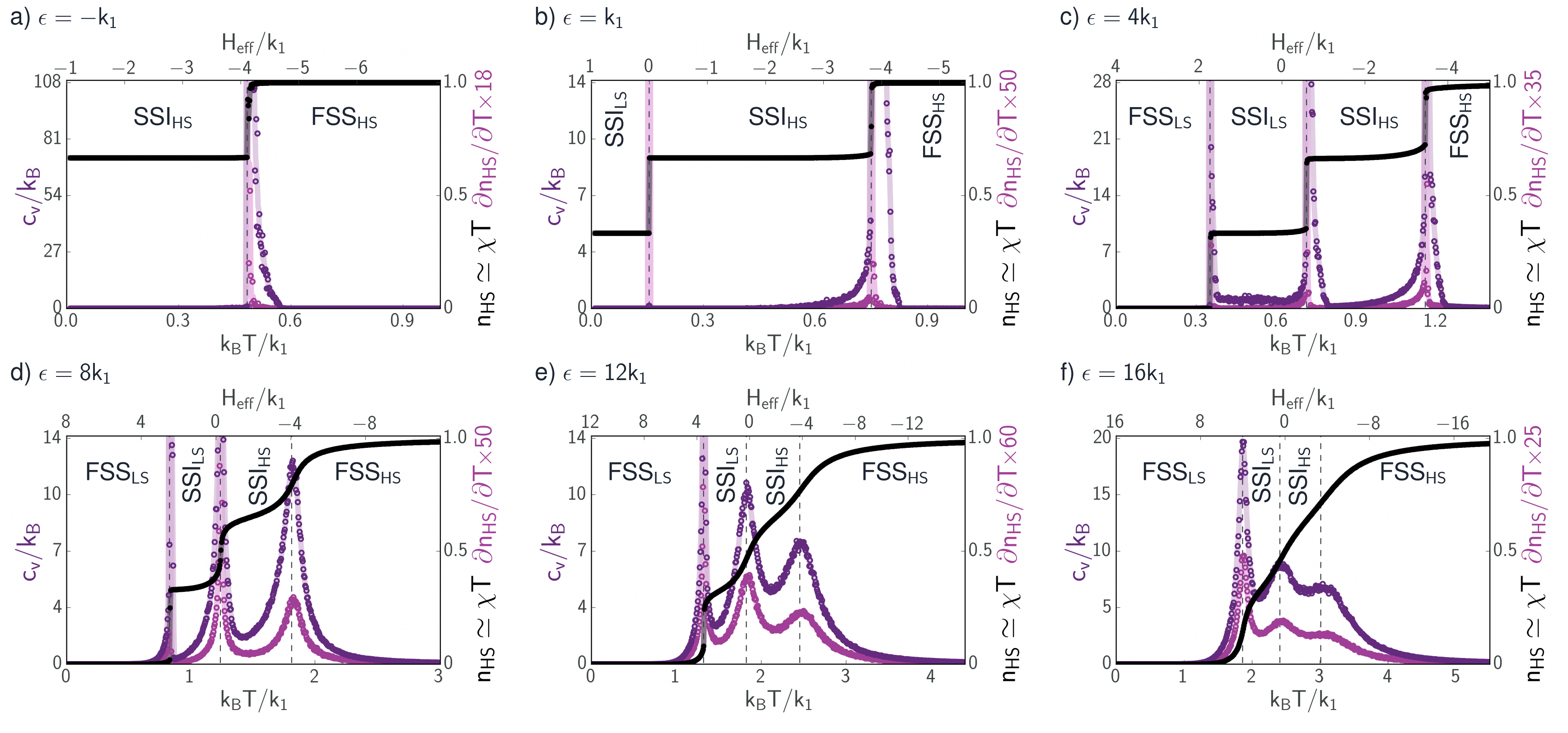}  
	\centering
	\caption{
		Thermodynamic signatures of spin-state ice. All panels show the heat capacity (left axes; purple), $c_V$; the fraction of HS metal centers (right axes; black), $n_{HS}=\frac12(1+m)\simeq\chi T$, where $\chi$ is the spin susceptibility, as the spins are weakly interacting; and $\partial n_{HS}/\partial T$ (right axes; pink). Both the data (points) and the SG filtered curve are shown for $n_{HS}$, observe that the difference is much smaller than the symbol size. The curves for $c_V$ and $\partial n_{HS}/\partial T$ are guides to the eye.
		a) For small negative enthalpy differences between the HS and LS states (here $\varepsilon=-k_1$) the ground state is SSI$_\text{HS}$ and there is a phase transition to a high temperature FSS$_\text{HS}$, which has majority HS metal centers. The FSS$_\text{HS}$ phase is adiabatically connected to the high temperature trivial phase.  
		b) For small positive enthalpy differences (here $\varepsilon=k_1$) the ground state is SSI$_\text{LS}$. As the temperature is raised there is a first order phase transition to the SSI$_\text{HS}$ phase, which shows significant ``melting'' (cf. Fig. \ref{Fig:snapshots}b) as witnessed by the increase in $n_{HS}$, followed by a transition to the trivial high temperature phase. 
		c) For larger positive enthalpy differences (here $\varepsilon=4k_1$) the FSS$_\text{LS}$ phase occurs at $T=0$. As the temperature is increased the system undergoes three phase transitions: first to  SSI$_\text{LS}$, second to  SSI$_\text{HS}$ and finally to  FSS$_\text{HS}$.
		As well as moving through the phase diagram (cf. Fig. \ref{fig:phase-dia}) large enthalpy differences also broaden the transitions: d) $\varepsilon=8k_1$;  e) $\varepsilon=12k_1$; and  f) $\varepsilon=16k_1$. For large enthalpy differences thermally induced SSI phases are quite narrow and the transitions are very broad. Thus, if only the fraction of HS states is calculated, or if only $\chi T$ is measured, the SSI phased could be dismissed as simply a broad single step HS to LS transition. However, $c_V$ and $\partial n_{HS}/\partial T$ both show clear signatures of the phase transition.
		%For different parameters we also find the classic SCO first order transitions directly between the FSS$_\text{LS}$ and FSS$_\text{HS}$/trivial phases. 
		In all panels the dashed vertical lines mark the phase transitions and $k_2=-0.139k_1$, $\Delta S=4k_B \ln5$,  $k_3=(3/4)k_2$.   
	}\label{Fig:thermo}
\end{figure*}

We make a uniform lattice approximation to the above model, i.e., we assume that for all nearest neighbors $r_{i,j}=x$ and that the topology of the lattice is not altered by changes in the spin-states. Minimizing with respect to $x$ and expanding the resulting Hamiltonian yields an effective Ising model in a longitudinal field 
\begin{eqnarray}
\mathcal{H}&=&\sum_{n=0}^{n_{max}}\mathcal{H}_n\notag\\
&=&\sum_{n=1}^{n_{max}} \sum_{\langle i,j\rangle_n} J_n \sigma_i \sigma_{j} - \frac{J_\infty}{N} \sum_{i,j}\sigma_i\sigma_j + H_\text{eff}\sum_i\sigma_i,~~ \label{model}
\end{eqnarray}
where $J_n=\alpha_n^2 k_n [R(1)-R(-1)]^2$, $\alpha_n$ is the distance to the $n^\text{th}$ nearest neighbor in units of $x$: $\{\alpha_n\}=\{1,\sqrt{3},2,\dots\}$, $J_\infty=4\sum_n J_n$ is a long-range strain interaction, which has equal strength between all sites regardless of their separation, $N$ is the number of lattice sites, and we have neglected a constant term.

Recently crystals of [Co$_2$Fe$_2$] complexes have been grown where these complexes lie at the vertices of a kagome lattice \cite{Sekine}. This system shows a three step SCO transition with macroscopic spin-state disorder in the low temperature phase.
%and a  three dimensional SCO framework on a kagome-like topology \cite{Neville}
Motivated by this and  recent synthetic progress towards kagome lattice SCO frameworks \cite{Suzanne-Jace}, we study model (\ref{model}) on the kagome lattice, Fig. \ref{Fig:model}e, with interactions up to third nearest neighbors. There are two distinct third nearest neighbors; we include the through-bond interaction, $k_3$, but neglect the through-space interaction $k_{3s}$, which one expects to be weaker. Generically, one expects $k_1>0$ because this distance will be close to minimum of the intermolecular interactions. However, one expects the longer range elastic constants $k_2$ and $k_3$ to be negative as long-range interactions generally fall off rapidly. This is shown explicitly for the Lennard-Jones potential in Fig. \ref{Fig:model}d. 

Our central result is that we find extended spin-state ice phases both at zero and non-zero temperatures, Figs. \ref{fig:phase-dia}--\ref{Fig:thermo}. 
At $T=0$ we have confirmed analytically %\cite{calcs} 
that there are no long-range ordered states with a unit cell of 48 or fewer sites that have lower energies than the spin-state ices in the relevant parts of the phase diagram, Fig. \ref{fig:phase-dia}. We find two spin-state ice phases: in the majority LS ice (SSI$_\text{LS}$; Fig \ref{Fig:rules}c) each triangle contains one HS and two LS metal centers; conversely the majority HS ice (SSI$_\text{HS}$; Fig \ref{Fig:rules}d) obeys the ice rule that each triangle contains two HS sites and one LS site. 
Thus, these phases are analogous to the three state ices of the kagome lattice Ising model in a field, which can be mapped to the honeycomb dimer model and have an extensive residual entropy $S=0.108 k_B$ associated with the Ising degrees of freedom  \cite{IsingKagome}. We also find ferro-spin-state (FSS) phases, characterized by a (large) pseudospin magnetization, $m=\sum_i\sigma_i/N$  and no local constraints.

However, the long-range strain interaction, $J_\infty$, leads to important differences between the zero-temperature phase diagram of model (\ref{model}) and the Ising model in a longitudinal field \cite{Udagawa}. Most importantly $J_\infty$ favors minimizing ($J_\infty<0$) or maximizing ($J_\infty>0$) the pseudospin magnetization. The short-range elastic interactions favor SSI for $k_2>-k_1/3$ and FSS for $k_2<-k_1/3$ (for $k_3=3k_2/4$). If $J_\infty$ is sufficiently strong it determines the ground state, with FSS realized for  $J_\infty$ large and positive and FSS$_\text{HS}$+FSS$_\text{LS}$ for  $J_\infty$ large and negative. For weak negative $J_\infty$ the long-range strain causes defects in the SSIs. The short range elastic interactions cause the defects to cluster and hence the coexistence of the various phases, as shown in Fig \ref{fig:phase-dia}.

\section{Monte Carlo simulations}

To investigate the finite temperature properties of  model (\ref{eq:Heff}) we carried out Monte Carlo simulations on a $N=40 \times 40 \times 3$ sites with periodic boundary conditions allowing single pseudospin flip, worm and loop moves \cite{carlos}. Additionally, we employ parallel tempering, switching between simulations with a Boltzmann probability. For each point we preform 5000 measurements on a 4800 site lattice after initializing for 4800 moves with 4800 moves between each measurement.

The calculation of the heat capacity, $c_V$, for this model has some subtleties. There are two contributions to the entropy, and thus the heat capacity, (i) the entropy associated with the individual metal ions, $S^{(1)}=m\Delta S/2$; and (ii) the many-body entropy associated with the configuration of Ising pseudospins, $S^{(N)}$. To calculate the single molecule term a Savitzky-Golay (SG) filter implementation \cite{SG} was used to fit $n_{HS}$. These fits are shown in Fig \ref{Fig:thermo}. The resulting analytic expressions were used to calculate  $\partial n_{HS}/\partial T$ and thence $c_V^{(1)}=T(\partial S^{(1)}/\partial T)=T\Delta S(\partial n_{HS}/\partial T)$.  The many-body contribution to heat capacity is evaluated from the fluctuations in the physical enthalpy, $E={\cal H} + T\Delta S\sum_i\sigma_i$, cf. Eq. (\ref{eq:Heff}),
\begin{eqnarray}
c_V^{(N)} = \frac{\langle E^2 \rangle - \langle E \rangle^2}{Nk_BT^2}. \label{eq:cVn}
\end{eqnarray}
Thus, the physical heat capacity is given by $c_V=c_V^{(1)}+c_V^{(N)}$.

\section{Spin-state ice at non-zero temperatures}

Snapshots from our Monte Carlo simulations  reveal states that obey the ice rules over a large temperature range, Fig. \ref{Fig:snapshots}a,b \cite{SI}. To confirm that the Monte Carlo calculations indeed find spin-state ices we have computed the pseudospin structure factor, 
\begin{equation}
{\mathcal S}_{\sigma\sigma}({\bm q})=\frac{1}{N^2}\sum_{ij}\left(\langle\sigma_i\sigma_j\rangle - m^2\right) e^{i{\bm q}.{\bm r}_{i,j}},
\label{eq:structure-factor-pseduospin}
\end{equation}
which we plot in Fig. \ref{Fig:snapshots}c. This clearly shows the pinch point singularities that are the signature of an ice state described by a divergence-less gauge field \cite{Henley,CastelnovoARCMP}. 

Pinch points are observed experimentally in neutron scattering experiments on spin ices \cite{Fennell}. However, the situation is more complicated for spin-state ice as the Ising degrees of freedoms are pseudospins and not directly amenable to neutron scattering. Given the large structural changes between the two spin-states it is possible that diffuse x-ray scattering  could directly measure the pseudospin structure factor. Alternatively, one could use an external magnetic field to align the spins on the HS sites. For a field in the $z$-direction one then has $S^z=\overline{S}+\sigma_i\delta S$, where $\overline{S}=(S^z_{HS}+S^z_{LS})/2$ and $\delta S=(S^z_{HS}-S^z_{LS})/2$. It follows straightforwardly that the spin structure factor,
\begin{eqnarray}
{\mathcal S}_{zz}({\bm q}) &=& \frac{1}{N^2}\sum_{ij}\left(\langle S^z_i S^z_j\rangle - m_{S}^2\right) e^{i{\bm q}.{\bm r}_{i,j}}\notag\\ 
&=& {\mathcal S}_{\sigma\sigma}({\bm q})\delta S^2, 
\label{eq:structure-factor-realspin}
\end{eqnarray} 
where $m_{S}=(1/N)\sum_i\langle S^z_i\rangle=\overline{S}+m\delta S$, is identical to the pseudospin structure factor except for a rescaling of the relative amplitude of the Bragg peak at the origin due to the ferri-spin-state (i.e., non-zero $m$) correlations in the spin-state ices.  As  the spin structure factor is directly measurable via neutron scattering this experiment could provide direct evidence for the existence of spin-state ices. Note that one expects that $\Delta S$ and thus $H_\text{eff}$ will depend on the applied magnetic field, so this requires that the SSI is stable in the magnetic field. 

Monte Carlo simulations also reveal local violations of the ice rules; two such excitations in SSI$_\text{LS}$ are marked in Fig. \ref{Fig:snapshots}a. The defects are a triangle with three LS sites ($\beta_{LS}$) and another with one LS and two HS sites ($\tau_{LS}$). Alternatively, one can view the SSI$_\text{LS}$ as long strings of LS sites; then the defects correspond to branching ($\beta_{LS}$) and terminating ($\tau_{LS}$) strings. To understand the properties of these excitations it is helpful to consider the system in a magnetic field in the $z$-direction that polarizes the spins. As each site belongs to two triangles, the SSI$_\text{LS}$ vacuum has a net spin $S^z_{\text{LSvac}}=\frac12(2S^z_\text{LS}+S^z_\text{HS})$ per triangle. Thus, a $\beta_{LS}$-excitation carries a net spin of $S^z_{\beta_\text{LS}}=\frac32 S^z_\text{LS}-S^z_{\text{LSvac}}=-\delta S$ and the net spin of an $\tau_{LS}$-excitation  is $S^z_{\tau_\text{LS}}=\frac12(S^z_\text{LS}+2S^z_\text{HS})-S^z_{\text{LSvac}}=\delta S$. 
Analogous excitations are found in the SSI$_\text{HS}$ phase, Fig. \ref{Fig:snapshots}b, with $S^z_{\text{HSvac}}=\frac12(2S^z_\text{HS}+S^z_\text{LS})$ per triangle; 
$S^z_{\beta_\text{HS}}=\delta S$ and 
$S^z_{\tau_\text{HS}}=-\delta S$. In contrast a single pseudospin flip is a $\pm2\delta S$ excitation. Depending on the number of $d$ electrons on the metal ion, it is possible for the fractionalized $\tau$ and $\beta$ excitations to carry either spin 1/2 or 1, Table \ref{table}.

To properly understand the physics of the defects a quantum description of the spin is required. For concreteness we  specialize to SSI$_\text{HS}$ and the most experimentally important case, $d^6$. The ideal SSI$_\text{HS}$ is characterized by strings of $S_{HS}=2$ ions separated by $S_{LS}=0$ ions. Although exchange interactions are small compared to elastic interactions that cause the SSI they are not negligible in the low $T$ quantum analysis of the spin. If the nearest neighbor exchange interactions  dominate, the spin Hamiltonian is simply the one-dimensional $S=2$ Heisenberg model along the HS strings. It is well known that this model has a Haldane gap \cite{Schollwock}. The $\tau_\text{HS}$ defects are terminating strings, and give rise to a fractionalized $S=1$ states localized at the ends of the strings \cite{Schollwock}. The $\beta_\text{HS}$ defects are branches in the strings and are topologically equivalent to a Y-junction, where again one finds a  $S=1$ fractionalized excitation localized at the defect \cite{Soos}. Thus, this quantum analysis is entirely consistent with the classical conclusion that the quasiparticles carry a fractionalized spin $\delta S=\Delta S_\text{SCO}/2$. For $d^4$, $d^5$, or $d^7$ $S_\text{LS}\ne0$, which makes the analysis considerably more complicated.

Spin-state transitions of a single ion on a  triangle where the ice rule is violated are low energy processes and allow defects to move. Explicitly this requires a majority-spin to minority-spin transition at a $\beta$ defect or a minority-spin to majority-spin transition at a $\tau$ defect. These processes change $\beta$ defects in $\tau$ defects and \textit{vice versa}. Thus, one immediately sees that the numbers of $\beta$ and $\tau$ defects are not conserved. However, there is a conserved topological charge associated with the motion of the excitations: $Q=\eta\delta S$, where $\eta=+1$ ($-1$) for $\bigtriangleup$ ($\bigtriangledown$) triangles \cite{Udagawa}.
At $T=0$ it only costs a finite energy to move at pair of excitations infinitely far apart. Hence, the fractionalized spin-$\delta S$ excitations are deconfined.

At non-zero temperatures the SSI vacuum can become unstable with respect to the spontaneous production of defects. As the spin-state ice is generically in an effective magnetic field ($H_\text{eff}\ne0$) the equilibrium states generically contain unequal numbers of $\beta$ and $\tau$ quasiparticles as the ice begins to melt. A snapshot of a melting ice, Fig. \ref{Fig:snapshots}b \cite{SI}, shows many defects but more branches than terminating strings, indicative of a greater fraction of HS ions  than for pure SSI$_\text{HS}$ ($n_{HS}=2/3$).

\begin{table}
	\centering
	\begin{tabular}{l|cccc} 
		& $d^4$ & $d^5$	& $d^6$ & $d^7$ \\ \hline 
		$S_{HS}$ 				& 2 	& 5/2  	& 2		& 3/2	\\  
		$S_{LS}$ 				& 1 	& 1/2	& 0		& 1/2 	\\  
		$\Delta S_\text{SCO}$  & 1 	& 2		& 2		& 1 	\\  
		$\delta S$ 			& 1/2 	& 1		& 1		& 1/2	\\  
	\end{tabular}
	\caption{
		Spin states relevant for octahedral complexes. SCO can occur for metal ions with $d^4$-$d^7$ electrons. $\Delta S_\text{SCO}=S_{HS}-S_{LS}=2\delta S$ is the magnitude in the change in spin when a single molecule undergoes SCO. $\delta S$ is the spin carried by the fractionalized quasiparticles.
	}
	\label{table}
\end{table}

While the spin-state ice rules that emerge from model (\ref{model}) are analogous to the spin ice rules for kagome ice in a field \cite{Wolf}, the model itself has important differences. The long-range strain plays a crucial role in determining which state is realized. First, let us consider the effects of $J_\infty$ at $T=0$ (cf. Fig. \ref{fig:phase-dia}). $J_\infty>0$ favors the maximization of the pseudospin magnetization and when the $J_\infty$ terms dominates one finds a FSS phase, with the spin-state selected by effective field, $H_\text{eff}=\varepsilon/2$. For weaker $J_\infty$ ($k_2$ and $k_3$ more negative) the short-range interactions dominate and stabilize spin-state ice states, again the majority spin-state is determined by $H_\text{eff}$. On the other hand $J_\infty<0$ favors equal numbers of high- and low-spin metal centers ($m=0$); this drives coexisting domains of the two polarizations in both the FSS and spin-state ice phases.

A second important difference emerges at finite temperature (cf. Fig. \ref{Fig:thermo}). $H_\text{eff}$ is temperature dependent and changes sign at $T=\varepsilon/\Delta S$, cf. Eq. (\ref{eq:Heff}). This can drive first order transitions from SSI$_\text{LS}$ to SSI$_\text{HS}$ or from FSS$_\text{LS}$ to FSS$_\text{HS}$, the latter corresponding to the classic spin-crossover transition. We also find parameter regimes where a SSI phase intervenes between the two FSS phases, Fig. \ref{Fig:thermo}c-f. At the highest temperatures studied the FSS$_\text{HS}$ crosses over to a trivial phase.  Thus, while scattering experiments could provide “smoking gun” evidence for spin-state ice, there are also clear signatures in thermodynamic probes, as shown in Fig. \ref{Fig:thermo}, which could provide an important means for the first identification of SSI phases experimentally.

The enthalpy difference between the HS and LS states ($\varepsilon=\varepsilon_{HS}-\varepsilon_{LS}$) has two important consequences for the finite temperature behavior of the system, Fig. \ref{Fig:thermo}.
Firstly, we have already seen that $\varepsilon$ plays an important role in determining the zero temperature physics (cf. Fig. \ref{fig:phase-dia}). This, in turn, helps determine the finite temperature behavior. For large positive $\varepsilon$ (Fig. \ref{Fig:thermo}c-f) we observe four phases. At the lowest temperatures an FSS$_\text{LS}$ phase that first gives way to a SSI$_\text{LS}$ phase as the temperature is raised. When the temperature raised sufficiently to change the sign of $H_\text{eff}$ [cf. Eq. (\ref{eq:Heff})] there is a transition from SSI$_\text{LS}$ to SSI$_\text{HS}$. Finally there is a transition from SSI$_\text{HS}$ to FSS$_\text{HS}$, which gradually becomes the trivial high temperature phase. Note that at high temperatures HS and LS metal ions are not equally likely because of their intrinsic entropy difference. The pseudospin magnetization $m\rightarrow\tanh(\Delta S/2k_B)$ as $T\rightarrow\infty$, which gives $n_{HS}\rightarrow625/626\simeq0.9984$ for $\Delta S=4k_B \ln5$, the value studied in Fig. \ref{Fig:thermo}. For small positive $\varepsilon$ (Fig. \ref{Fig:thermo}b) the SSI$_\text{LS}$ phase is found at $T=0$ and the FSS$_\text{LS}$ phase is not realized at any temperature, but otherwise the finite temperature physics is quite similar to large  positive $\varepsilon$, with the SSI$_\text{HS}$ and FSS$_\text{HS}$/trivial phases realized as the temperature is raised. For small negative $\varepsilon$ (Fig. \ref{Fig:thermo}a) one has the SSI$_\text{HS}$ phase at zero temperature. No majority LS phases are found at any temperature. Nevertheless, the finite temperature physics is similar to that of for positive $\varepsilon$, but with the low temperature majority LS phases removed. For large negative $\varepsilon$ (not shown) the FSS$_\text{HS}$ phases occurs at all temperatures. This can all be understood straightforwardly as the $\varepsilon$ sets the minimum $H_\text{eff}$, cf. Eq. (\ref{eq:Heff}).

Secondly, the breadth of the transition is highly dependent on enthalpy difference between the HS and LS states. For small $|\varepsilon|$ we see sharp, well defined signatures of the phase transitions via both peaks in the heat capacity and steps in the fraction of HS ions. As $\varepsilon$  increases these features become increasingly washed out and crossover-like. The magnetic susceptibility, $\chi$, is the most commonly measured property of SCO materials: because of the extremely weak spin interactions typically found in these materials, $n_{HS}\propto\chi T$. For the largest $\varepsilon$ studied (Fig. \ref{Fig:thermo}f) the SSI phases are confined to a narrow region around $H_\text{eff}=0$ and the changes in $n_{HS}$ occur rather slowly. 
Thus, if only $\chi T$ were measured SSI physics could easily be dismissed as a broad single step SCO between FFS$_\text{LS}$ and FFS$_\text{HS}$, which are rather common. On the other hand, there are clear maxima in the specific heats even at $\varepsilon=16k_1$, showing that clear thermodynamic signatures of phase transitions or crossovers into, out of, and between the SSI phases remain. These maxima are also found in $\partial n_{HS}/\partial T$, which is a more routine measurement in the spin crossover community. 
It is interesting to note that this is not just a straightforward consequence of $c_V^{(1)}\propto\partial n_{HS}/\partial T$. 
Rather, we find that $c_V^{(N)}\gg c_V^{(1)}$ in all cases studied: even near the phase transitions. Thus, the critical fluctuations present in both the single-body and many-body contributions to the entropy.

\section{Outlook}

Clear evidence of spin-state disorder has recently been found from x-ray crystallographic studies of the intermediate temperature phase of [Co$_2$Fe$_2$] complexes on kagome lattices \cite{Sekine}. Our works suggests that spin-state ice physics is at play. However, in this material the three sublattices of the kagome lattice sit at three crystallographically inequivalent sites. Thus, one should expect, at least, different $\varepsilon$ on the different sublattices, which complicates the emergence of ice physics.

Natural questions arising for our work include: what is the quantum mechanical ground state of the elastic Hamiltonian? And how large are the quantum effects? Both the spins and the harmonic oscillators are quantum objects. Based on the analogy to spin ices, one might expect a quantum spin-state liquid. As quantum effects are large for phonons, one might expect the quantum effects to be significant.  Furthermore, as light and spin-orbit coupling can drive transitions between LS and HS (and vice versa) \cite{Kahn,Goodwin,Halcrow}, they may act as ``transverse fields'' in the Ising-like model. 

\acknowledgements

We thank Cameron Kepert,  Yaroslav Kharkov, Grace Morgan, Ross McKenzie, Suzanne Neville, Blake Peterson, Nic Shannon and Xiuwen Zhou for helpful conversations. Ross McKenzie  provided the argument summarized in Fig. \ref{Fig:model}d. This work was funded by the Australia Research Council (FT130100161) and an Australian Government Research Training Program Scholarship.

\renewcommand\thefigure{S\arabic{figure}}
\setcounter{figure}{0}  

\section*{Supplementary Information}

%\begin{figure}[p]
%	\includegraphics[width=0.9\columnwidth]{1_yeahbuddy}  
%	\centering
%	\caption{Density of HS ions, $n_{HS}$, and specific heat, $c_v$ as a function of temperature for $\varepsilon=k_1$, $k_2=-0.139k_1$, $k_3 = 3k_2/4$.   
%	}
%\end{figure}

\begin{figure*}
	\includegraphics[width=\textwidth]{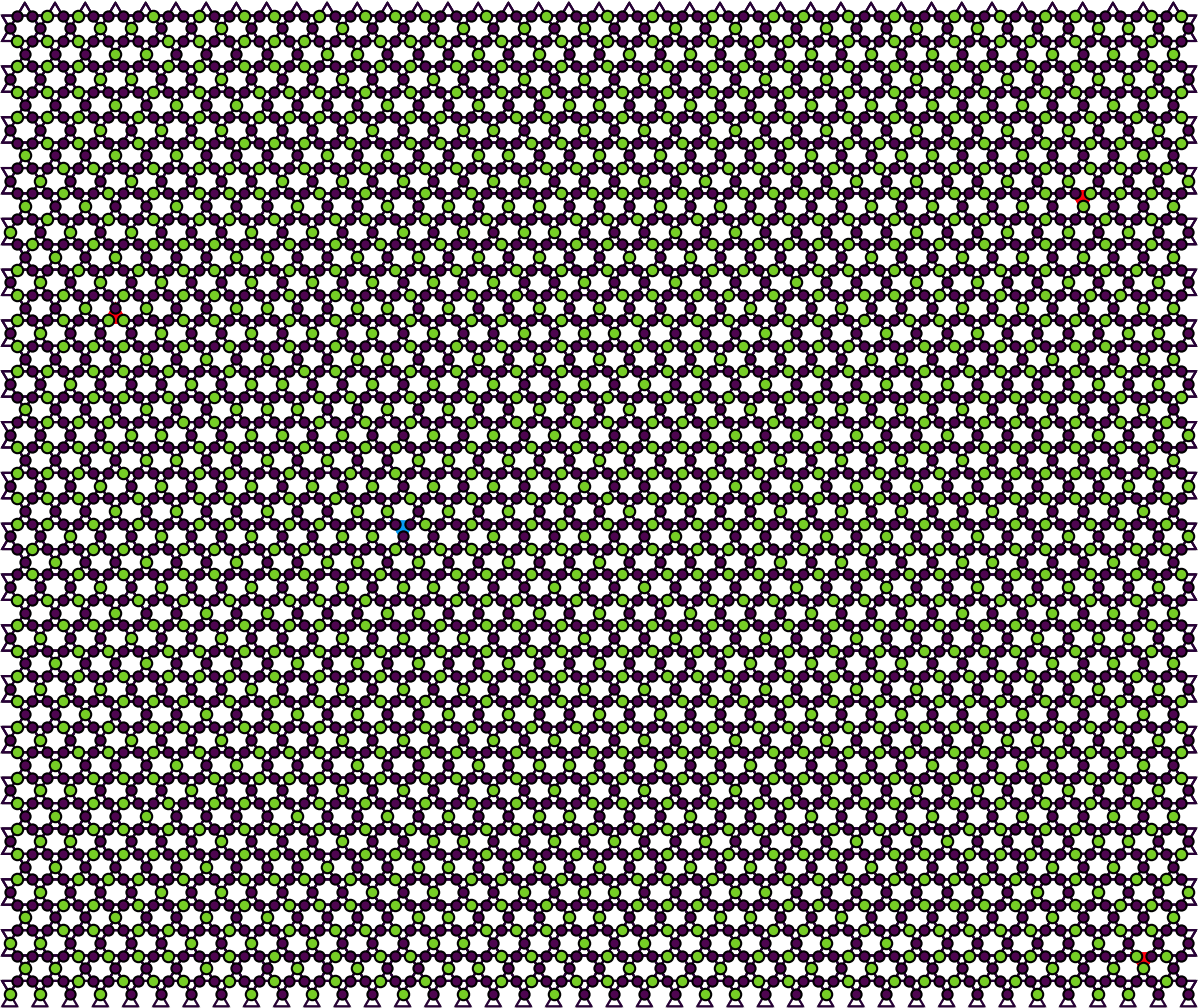}  
	\centering
	\caption{Snapshot of the full lattice from which Fig. \ref{Fig:snapshots}a is taken. 
	}
\end{figure*}

\begin{figure*}
	\includegraphics[width=\textwidth]{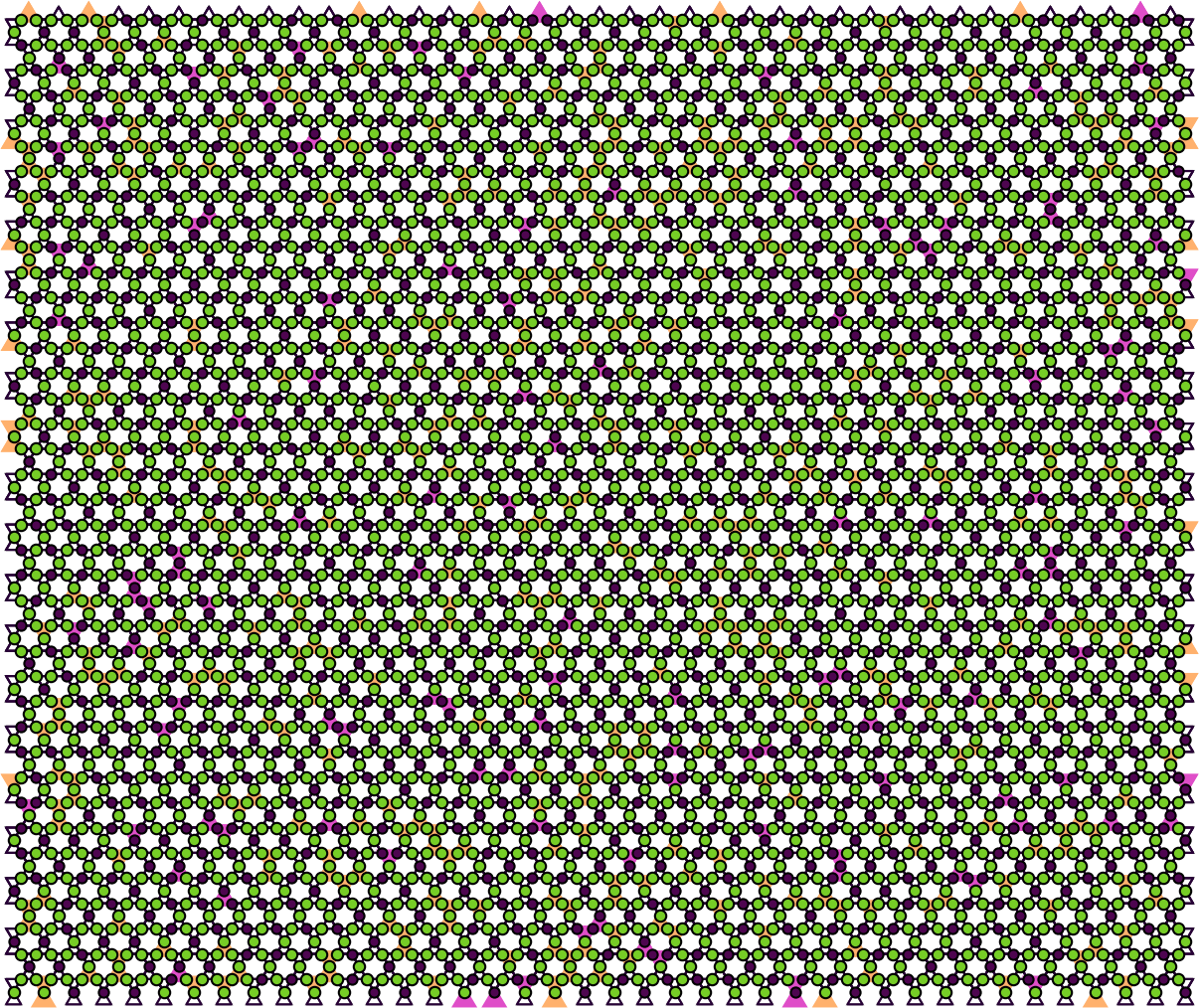}  
	\centering
	\caption{Snapshot of the full lattice from which Fig. \ref{Fig:snapshots}b is taken.
	}
\end{figure*}

\end{document}